\documentclass[aps,prd,twocolumn,superscriptaddress,nofootinbib,showpacs,10pt]{revtex4-1}
\usepackage[utf8]{inputenc}
\usepackage[english]{babel}
\usepackage{amsmath}
\usepackage{amsfonts}
\usepackage{amssymb}
\usepackage{graphics}
\usepackage{graphicx}
\usepackage[left=1.7cm,right=1.7cm,top=2.0cm,bottom=2.0cm]{geometry}
\usepackage[usenames,dvipsnames,svgnames]{xcolor}  
\usepackage{hyperref}   
\hypersetup{colorlinks=true,linkcolor=beamer@PRD, citecolor=beamer@PRD}
\definecolor{beamer@PRD}{RGB}{46,48,146}

\newcommand\myref[1]{\textcolor{beamer@PRD}{(}\ref{#1}\textcolor{beamer@PRD}{)}}

\begin{document}
\title{Generalized squeezed states}
\author{Kevin Zelaya}\email{kdzelaya@fis.cinvestav.mx}
\affiliation{Physics Department, Cinvestav, AP 14-740, 07000 M\'exico City, Mexico}
\affiliation{Centre de Recherches Math\'ematiques, Universit\'e de Montr\'eal, Montr\'eal--H3C 3J7, QC, Canada}
\author{Sanjib Dey}\email{dey@iisermohali.ac.in}\email{sanjibdey4@gmail.com}
\affiliation{Department of Physical Sciences, Indian Institute of Science Education and Research Mohali, \\ Sector 81, SAS Nagar, Manauli 140306, India}
\author{V\'eronique Hussin}\email{veronique.hussin@umontreal.ca}
\affiliation{Centre de Recherches Math\'ematiques, Universit\'e de Montr\'eal, Montr\'eal--H3C 3J7, QC, Canada}
\affiliation{D\'epartment de Math\'ematiques et de Statistique, Universit\'e de Montr\'eal, \\ Montr\'eal--H3C 3J7, QC, Canada} 
\begin{abstract}
Squeezed states are one of the most useful quantum optical models having various applications in different areas, especially in quantum information processing. Generalized squeezed states are even more interesting since, sometimes, they provide additional degrees of freedom in the system. However, they are very difficult to construct and, therefore, people explore such states for individual setting and, thus, a generic analytical expression for generalized squeezed states is yet inadequate in the literature. In this article, we propose a method for the generalization of such states, which can be utilized to construct the squeezed states for any kind of quantum models. Our protocol works accurately for the case of the trigonometric Rosen-Morse potential, which we have considered as an example. Presumably, the scheme should also work for any other quantum mechanical model. In order to verify our results, we have studied the nonclassicality of the given system using several standard mechanisms. Among them, the Wigner function turns out to be the most challenging from the computational point of view. We, thus, also explore a generalization of the Wigner function and indicate how to compute it for a general system like the trigonometric Rosen-Morse potential with a reduced computation time.
\end{abstract}

\pacs{}

\maketitle
\section{Introduction} \label{sec1}
Squeezed states are interesting quantum optical systems exhibiting nonclassical properties \cite{Walls,Loudon_Knight,Teich_Saleh}. They produce less noise in optical communication than a vacuum state. This is why squeezed light has various applications in different areas including optical communications \cite{Yamamoto}, optical measurement \cite{Caves}, detection of gravitational waves \cite{Vahlbruch,Chua}, universal quantum computing \cite{Menicucci}, dense coding  \cite{MasashiBan}, etc. Squeezed states are also utilized in quantum metrology not only to improve the quantum metrology technique itself \cite{Dowling,Giovannetti,Riedel}, but also to increase the sensitivity of gravitational wave detectors \cite{Vahlbruch1,Schnabel,Anisimov}, especially the LIGO \cite{Aasi}. Moreover, squeezed light serves as a primary resource in continuous variable quantum information processing and, is utilized to distribute secret keys in quantum cryptography \cite{Hillery}. For an extensive list of applications one may refer, for instance \cite{Braunstein,Andersen}.

The primitive idea of squeezing follows from the Heisenberg uncertainty principle. It is well-known that the coherent states of light minimize the uncertainty relation $\Delta x\Delta p=1/2$, with both of the quadrature uncertainties being identical to each other $\Delta x=\Delta p=\sqrt{1/2}~(\hbar =1)$. Consequently, the uncertainty region of a coherent state can be represented by a circle in the optical phase space. However, for some states the uncertainty circle may be squeezed in one quadrature and elongated correspondingly in the other so that the uncertainty circle is deformed to form an ellipse and, the corresponding states are often familiar as squeezed states. Although, the uncertainty relation does not necessarily have to be minimized in the latter case, however, there are rare examples for squeezed states with minimum uncertainties \cite{Yamamoto_etal,Dey_Fring_Hussin}, which are popular as \textit{ideal squeezed states}. Nevertheless, so far we have discussed a particular type of squeezed states, namely the \textit{quadrature squeezed} states, which are defined as the states whose standard deviation in one quadrature is less than that of the coherent states or a vacuum state. Squeezing can also occur in photon number distribution, and a state is said to be \textit{number squeezed} if the photon number uncertainty of the corresponding states becomes lower than that of the coherent states. However, physically both of the scenarios refer to the notion of nonclassicality.

Coherent states are not nonclassical, in fact, they are the most classical analogue of quantum systems. But, the statement is true only for the coherent states of the harmonic oscillator, which are sometimes referred to the Glauber coherent states. However, there are various generalized coherent states \cite{Barut,Stoler,Arik_Coon,Perelomov_Book,Manko_Marmo_Sudarshan_Zaccaria,Ali_Antoine_Gazeau,Sivakumar,Dodonov} for which the quadrature and/or number squeezing occur and, thus, they are nonclassical. People sometimes refer these types of coherent states or any other nonclassical states to be \textit{squeezed states} for which the quadrature and/or number squeezing occurs. It should be noted that in this article we do not refer these types of nonclassical states to be squeezed states, rather, we talk about a particular class of states as defined in Sec. \ref{sec2}, which are constructed in such a way that the quadrature squeezing is inherited to them by construction and, thus, they are always nonclassical.

Generalization of different quantum optical systems provides a deeper understanding, since sometimes it brings additional degrees of freedom in the system so that they can be applied more efficiently to physical models \cite{Dodonov}. The essence of the generalization of several nonclassical states; such as, cat states \cite{Xia_Guo,Filho_Vogel,Mancini,Dey}, photon-added coherent states \cite{Agarwal_Tara,Duc_Noh,Safaeian,Dey_Hussin_Photon}, pair-coherent states \cite{Agarwal_Biswas}, binomial states \cite{Lee}, etc., have been explored in various contexts. Squeezed states are probably the most well-behaved nonclassical states which can be prepared more systematically and elegantly in the laboratory \cite{Wineland}. However, as per our knowledge, there is no particular form of the generalized squeezed states available in the literature, because they are extremely difficult to construct. In this article, we propose a generalization to such states followed by an example of a general system, viz., the trigonometric Rosen-Morse potential, where we apply our proposal directly. The reason behind choosing the Rosen-Morse potential for our analysis is mainly because it is a widely used model in optics, but mostly because the squeezed states for such model have not been studied notably. Therefore, we have the opportunity to explore in a two-fold way. Firstly, the generalization of the squeezed states is verified via an example through a popular model. Secondly, at the same time, we can shed light on the behavior of the Rosen-Morse squeezed states, which is inadequate in the literature. It should be noted that there are many articles available, for instance \cite{Satyanarayana,Braunstein_McLachlan,Ma_Rhodes,Katriel_Solomon,Lo_Sollie, Nieto_Truax_PRL,Seshadri_Lakshmibala,Trifonov,Hong_Hai-Ling,Alvarez_Hussin,Kwek_Kiang, Obada_Al-Kader,Shchukin}, entitled by ``generalized squeezed states", however, they contain either the ``generalized coherent states" having the squeezing properties or, they are any other type of nonclassical states whose quadrature and/or photon number is/are squeezed. However, according to our knowledge there is no trace of the generalization of the particular state that we discuss in the following section.   

In Sec.~\ref{sec2}, we discuss the detailed procedure for the generalization of the squeezed states by introducing a set of generalized ladder operators followed by an explicit analytic solution of the generalized squeezed states. In Sec.~\ref{sec3}, we apply the obtained general solution to a particular type of model, namely the Rosen-Morse potential. Sec.~\ref{sec4} is composed of the analysis of nonclassicality and squeezing properties of the given example by means of the analysis of quadrature squeezing, sub-Poissonian photon statistics and Wigner distribution function. Finally, our concluding remarks are stated in Sec.~\ref{sec5}.
\section{Generalization}\label{sec2}
Squeezed states for harmonic oscillator $\vert \alpha,\delta\rangle_{\text{ho}}$ are constructed by operating the  displacement operator $D(\alpha)=exp(\alpha a^\dagger-\alpha^\ast a)$ on the squeezed vacuum $S(\delta)\vert 0\rangle$ \cite{Nieto_Truax_PRL}
\begin{equation}\label{SqOperator}
\vert \alpha,\delta\rangle_{\text{ho}}=D(\alpha)S(\delta)\vert 0\rangle, \quad S(\delta)=e^{\frac{1}{2}(\delta a^\dagger a^\dagger-\delta^\ast aa)},
\end{equation}
with $\alpha,\delta\in \mathbb{C}$ being displacement and squeezing parameters, respectively. Alternatively they can be formulated by performing the Holstein-Primakoff/Bogoliubov transformation on $S(\delta)$ arising from the solution of the following eigenvalue equation \cite{Fu_Sasaki,Nieto_Truax,Alvarez_Hussin}
\begin{equation}\label{eigen}
(a+\xi a^\dagger) \vert \alpha,\xi\rangle_{\text{ho}}=\alpha \vert \alpha,\xi\rangle_{\text{ho}}, ~\xi=\frac{\delta}{|\delta|}\tanh(|\delta|), ~|\xi|<1,
\end{equation}
which reduces to the coherent states for $\xi=0$. Generalization are usually carried out \cite{Alvarez_Hussin,Angelova_Hertz_Hussin} by replacing the usual ladder operators $a, a^\dagger$ by the generalized ladder operators $A, A^\dagger$ in \myref{eigen}
\begin{equation} \label{GenLad}
A^\dagger \vert n\rangle = \sqrt{k(n+1)} \vert n+1\rangle, \quad A \vert n\rangle = \sqrt{k(n)} \vert n-1\rangle,
\end{equation}  
with $k(n)$ being an operator-valued function of the number operator $n=a^\dagger a$ associated with generalized models. The generalized ladder operators \myref{GenLad} have been introduced long time back \cite{Manko_Marmo_Sudarshan_Zaccaria,Sivakumar,Filho_Vogel,Mancini} in order to generalize various quantum optical models, especially coherent and cat states. These generalizations are mostly familiar as nonlinear generalization and they are well-established and widely accepted in the community. The existence of such states have also been found in many experiments using Kerr type nonlinearity and nonlinear cavity \cite{Wang_Goorskey_Xiao,Gambetta,Yan_Zhu_Li}. For further information in this regard one may follow some review articles in the context \cite{Dodonov,Dey_Review}. Note that, the generalized ladder operators \myref{GenLad} are given in such a form that $A^\dagger A$ behaves as the number operator of the generalized system and, therefore, the function $k(n)$ can be associated with the eigenvalues $e_n$ of the model as follows
\begin{equation}
A^\dagger A|n\rangle=k(n)|n\rangle , \qquad k(n)\sim e_n,
\end{equation}
which holds in general for the function $k(n)$. The appearance of additional constant terms in the eigenvalues can be realized by rescaling the composite system of $A$ and $A^\dagger$ correspondingly. Therefore, by computing the eigenvalues of the system, one can construct the function $k(n)$ and, thus, various quantum optical states by using the nonlinear generalization procedure that is discussed above.  Nevertheless, in order to solve the eigenvalue equation \myref{eigen} in the generalized scenario, let us first expand the squeezed states in Fock basis
\begin{equation}\label{Expansion}
\vert \alpha,\xi\rangle=\frac{1}{\mathcal{N}(\alpha,\xi)}\displaystyle\sum_{n=0}^\infty \frac{\mathcal{J}(\alpha,\xi,n)}{\sqrt{k(n)!}}\vert n \rangle~,
\end{equation} 
where $k(n)!=\prod_{i=1}^n k(i)$ and $k(0)=1$. By inserting \myref{Expansion} into \myref{eigen} replaced by the generalized ladder operators \myref{GenLad}, we end up with a three term recurrence relation
\begin{equation}\label{recurrence}
\mathcal{J}(\alpha,\xi,n+1)=\alpha~\mathcal{J}(\alpha,\xi,n)-\xi ~k(n)~ \mathcal{J}(\alpha,\xi,n-1),
\end{equation}
with $\mathcal{J}(\alpha,\xi,0)=1$ and $\mathcal{J}(\alpha,\xi,1)=\alpha$, which when solved one obtains the explicit form of the squeezed states \cite{Angelova_Hertz_Hussin}. But, the recurrence relation \myref{recurrence} is extremely difficult to solve for general $k(n)$, in fact, there are only few examples where it has been possible, that too for some particular models  \cite{Angelova_Hertz_Hussin,Dey_Hussin}. Notice that, Eq.~\myref{recurrence} reduces to a simple form for $\xi=0$, which is the case of coherent states and, indeed, the corresponding solution leads to the nonlinear coherent states \cite{Filho_Vogel,Manko_Marmo_Solimeno_Zaccaria,Sivakumar}
\begin{equation}\label{NonLinear}
\vert \alpha\rangle=\frac{1}{\mathcal{N}(\alpha)}\displaystyle\sum_{n=0}^\infty \frac{\alpha^n}{\sqrt{k(n)!}}\vert n \rangle~,
\end{equation}
which are the generalized version of the Glauber coherent states. Furthermore, in order to obtain the standard form of the harmonic oscillator squeezed states, we must consider $k(n)=n$ and, with this the recurrence relation \myref{recurrence} is solved to obtain
\begin{equation}\label{SqStateHO}
\vert \alpha,\xi\rangle_{\text{ho}}^{}=\frac{1}{\mathcal{N}(\alpha,\xi)}\displaystyle\sum_{n=0}^\infty \frac{1}{\sqrt{n!}}\left(\frac{\xi}{2}\right)^{n/2}\mathcal{H}_n(\frac{\alpha}{\sqrt{2\xi}})\vert n \rangle~,
\end{equation}
where $\mathcal{H}_n(\alpha)$ denote the complex Hermite polynomials. We intend to provide a general solution applicable for any model, which is obtained by the general solution of \myref{recurrence} as follows
\begin{eqnarray}\label{GenSqState}
\mathcal{J}(\alpha,\xi,n)=\displaystyle\sum_{m=0}^{[n/2]}(-\xi)^m\alpha^{n-2m}g(n,m),
\end{eqnarray} 
with
\begin{equation}\label{gnm}
g(n,m)=\displaystyle\sum_{j_1=1}^{(n-2m+1)}\displaystyle\sum_{j_2=j_1+2}^{(n-2m+3)}\displaystyle\sum_{j_3=j_2+2}^{(n-2m+5)}\cdot\cdot\cdot\displaystyle\sum_{j_m=j_{m-1}+2}^{(n-1)}\mu,
\end{equation}
where $\mu=k(j_1)k(j_2)\cdot\cdot\cdot k(j_m)$ and $g(n,0)=1$. The notation $[n]$ in \myref{GenSqState} represents the Floor function, whose output is an integer less than or equal to the corresponding real number $n$. The solution \myref{GenSqState} is achieved initially by computing the first few terms of the recurrence relation \myref{recurrence} with initial conditions $\mathcal{J}(\alpha,\xi,0)=1$ and $\mathcal{J}(\alpha,\xi,1)=\alpha$. Thereafter, a relatively closed form of the series is acquired by the trial and error method, which is never an elegant way to solve a recurrence relation. However, once the solution is obtained it is straightforward to verify whether the solution satisfies the recurrence relation. For the detailed proof one may refer to the Appendix. Eq.~\myref{GenSqState} along with \myref{gnm}, thus, provides a powerful solution of \myref{recurrence}, which can be employed to obtain the squeezed states for any model corresponding to the known eigenvalues $k(n)$. Let us now study how the general solution \myref{GenSqState} works for a given system and verify their nonclassical properties. The main aim is to extract the factor $k(n)$ from the expression of the eigenvalues of the given model and apply it to the general solution \myref{GenSqState} to obtain the squeezed states corresponding to it.
\section{Example: Rosen-Morse potential}\label{sec3}  
The trigonometric Rosen-Morse potential \cite{Rosen_Morse,Compean}
\begin{equation}
V(x)=-2b\cot x +d(d+1)\csc^2x, \quad \forall x\in [0,\pi],
\end{equation}
is one of the well-studied models and the corresponding energy eigenvalues are well-known \cite{Compean}
\begin{equation} \label{En}
E_n=(n+d+1)^2-\frac{b^2}{(n+d+1)^2}, \quad n=0,1,2,....,
\end{equation}
with $b,d$ being positive constants. While the model can be solved by any standard method available in the literature, we consider a particular procedure suitable for our purpose. We can construct the generalized ladder operators $A, A^\dagger$ defined by \myref{GenLad} along with 
\begin{equation}\label{kn}
k(n)=E_n-E_0=n(n+2d+2)\left[1+\frac{b^2}{(d+1)^2(n+d+1)^2}\right],
\end{equation}
so that the operator $H=A^\dagger A+E_0$ reproduces the spectrum \myref{En} in the Fock basis, with $E_0$ being the  ground state energy. Thus, it becomes straightforward to compute the squeezed states for the model by utilizing \myref{GenSqState} along with the $k(n)$ \myref{kn}. Let us now analyze their nonclassical behavior to ensure that the constructed squeezed states are well-behaved.
\section{Nonclassical properties}\label{sec4} 
The behavior of the coherent states being analogous to that of the classical objects, they are often familiar as classical like states, although originally they are the superposition of large number of quantum states. According to the convention of Glauber and Sudarshan \cite{Glauber,Sudarshan}, the quantum states which are less classical than the coherent states can be called nonclassical states. Technically they can be realized in terms of the Glauber-Sudarshan's $P$-function for arbitrary density matrices
\begin{figure*}
\centering \includegraphics[scale=0.265]{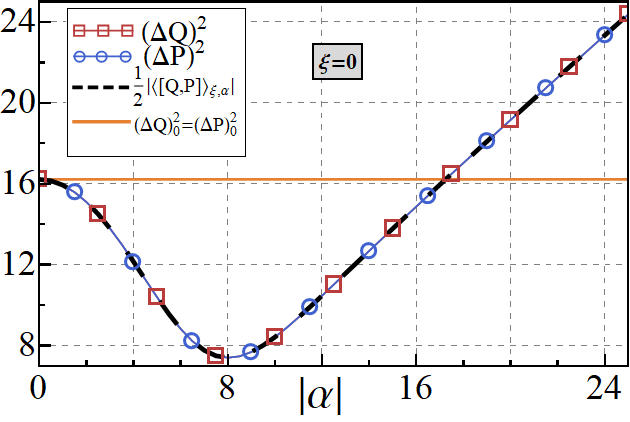} 
\includegraphics[scale=0.265]{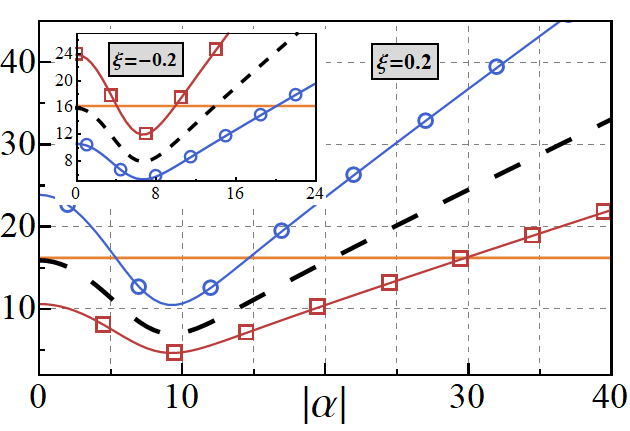}
\includegraphics[scale=0.265]{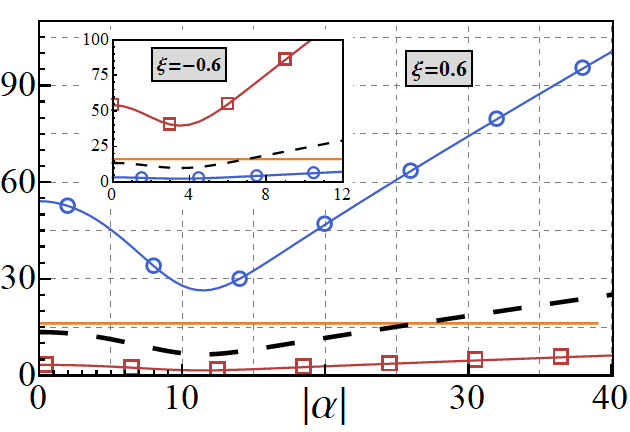}
\caption{Quadrature squeezing for the Rosen-Morse squeezed states.}
\label{fig1}
\end{figure*}
\begin{equation}
\rho=\int \text{d\,Re}\alpha~\text{d\,Im}\alpha~P(\alpha)|\alpha\rangle\langle\alpha|,
\end{equation}
with $\int\text{d\,Re}\alpha~\text{d\,Im}\alpha~P(\alpha)=1$. In case of coherent states, the weight function $P(\alpha)$ is a delta function and, thus, the function $P(\alpha)$ represents a probability density. On the other hand, the states for which the $P$-distribution fails to be a probability density are called nonclassical states. More precisely, if the singularities of the $P$-functions are either of types stronger than those of the delta functions (e.g. derivatives of delta function) or they are negative, the corresponding states have no classical analogue. For further details in this regard one may refer, for instance \cite{Dodonov,Dey_Review}. The convention of nonclassicality in case of our generalized systems is same as described above. The reason is that the generalization is carried out in such a way that the generalized ladder operators $A$ and $A^\dagger$ \myref{GenLad} operate on the Fock states $|n\rangle$ producing the generalized coherent states \myref{NonLinear} as well as the squeezed states \myref{GenSqState} in the Fock basis. Therefore, the whole construction is in the optical basis dealing with the optical photon number $a^\dagger a$. The only thing that remains is to understand the role of the extra nonlinear factor $k(n)$ that appears due to the generalization.  Note that, the signature of $k(n)$  underlies within the generalized ladder operators $A^\dagger$ and $A$, which constitute the commutation relation $[Q,P]$. And this commutation relation $[Q,P]$ is the driving factor in the RHS of the generalized uncertainty relation \myref{GenUncert} as given in the next section. Therefore, only a slight modification in the statement of the quadrature squeezing may be sufficient to understand the quadrature squeezing properties of our system. In the studies of photon number squeezing and Wigner function, we may not require any further modification to the standard definition, since the generalized squeezed states are constructed in the standard optical basis. However, a proper generalization of the Mandel parameter could provide further deeper understanding on the photon number squeezing of our systems, but it is not a trivial work. At least, to our knowledge, a proper generalization of such thing is not yet adequate in the literature. We have provided a more rigorous and technical analysis in this regard in the respective places in the subsequent sections. Let us now analyze some standard techniques to realize the nonclassical properties of our system.
\subsection{Quadrature squeezing}
The quadrature operators for the generalized system as defined by $Q=(A+A^\dagger)/\sqrt{2}$ and $P=(A-A^\dagger)/\sqrt{2}i$ obey the generalized  uncertainty relation
\begin{equation}\label{GenUncert}
\Delta Q\Delta P\geq \frac{1}{2}\big\vert\langle\alpha,\xi\vert[Q,P]\vert\alpha,\xi\rangle\big\vert.
\end{equation}
The relation \myref{GenUncert} holds for any arbitrary state $|\psi\rangle$ in general, however, since our analysis is based on the squeezed states $\vert\alpha,\xi\rangle$, we have written it in a particular form in terms of the squeezed states $\vert\alpha,\xi\rangle$. For the case of the vacuum state $|0\rangle$, the RHS of \myref{GenUncert} and square of each of the variances become equal each other, i.e. $(\Delta Q)^2=(\Delta P)^2=\frac{1}{2}\big\vert\langle 0\vert[Q,P]\vert 0\rangle\big\vert=k(1)/2$, and the same identity holds for the generalized coherent states \myref{NonLinear} also. Note that, since $[Q,P]$ is not in general proportional to the identity, the RHS of \myref{GenUncert} strongly depends on the state that is used to compute the expectation values. So, it can not be guaranteed that the variances associated with the generalized squeezed states are necessarily below to those of the vacuum state $|0\rangle$, as it occurs in the case of harmonic oscillator. Therefore, in our case, the quadrature squeezing does not correspond to the case when the variance of any of the quadratures becomes lower than the square root of the RHS of \myref{GenUncert} for the vacuum state $|0\rangle$. But, it corresponds to that of the particular state that is being studied, which is $|\alpha,\xi\rangle$ in our case. The behavior is depicted in Fig.~\ref{fig1}, where we plot the variance of the quadratures for the squeezed states $\Delta Q,\Delta P$ as well as for the vacuum state $(\Delta Q)_0,(\Delta P)_0$ along with the RHS of \myref{GenUncert} for different values of $\xi$. Panel (a) corresponds to the case of coherent states, and we notice that $\Delta Q, \Delta P$ and RHS of \myref{GenUncert} coincide. It means that the states belong to the category of \textit{intelligent states} \cite{Aragone} implying no quadrature squeezing with the variance of both of the quadratures being identical to each other. More interesting behavior is observed in the remaining two panels, in panel (b), the $Q$ quadrature is squeezed for $\xi=0.2$, whereas the $P$ quadrature squeezes for $\xi=-0.2$ (see the subpanel in panel (b)). The same holds true for panel (c), however, in this case the squeezing has increased compared to (b), as expected. Interestingly, in both of the cases the uncertainty relation \myref{GenUncert} is minimized. This is actually more exciting, since this is found very rarely in the literature, and is familiar as \textit{ideal squeezed states} \cite{Yamamoto_etal,Dey_Fring_Hussin}. The other interesting observation from the panel (b) and (c) is that the variances for the generalized squeezed states $\Delta Q, \Delta P$ are not always below to those of the vacuum state $(\Delta Q)_0,(\Delta P)_0$, however, they are always below the generalized uncertainty minimum $\frac{1}{2}|\langle [Q,P]\rangle|_{|\alpha,\xi\rangle}$. It confirms that the information of nonclassicality in case of the generalized squeezed states must be extracted from the minimum of the generalized uncertainty relation.
\subsection{Photon number squeezing}
\begin{figure}
\centering \includegraphics[scale=0.33]{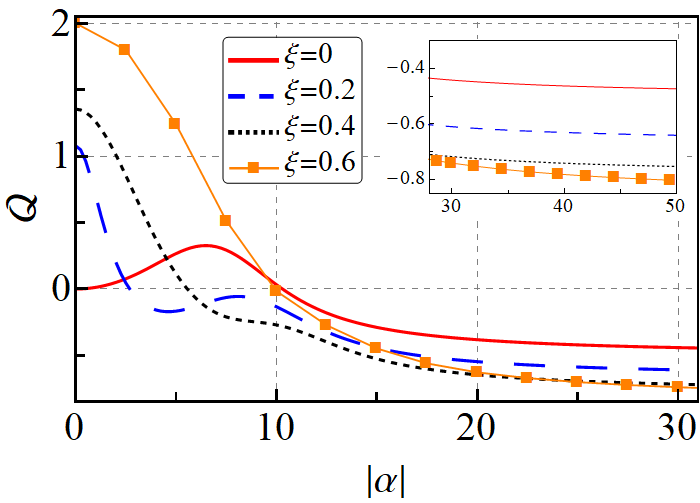}
\caption{Mandel parameter $\mathcal{Q}$ \myref{MandelQ} for the Rosen-Morse squeezed states for different values of $\xi$.}
\label{fig2}
\end{figure}

Photon number squeezing is another excellent method by which the nonclassicality of a state can be tested. A photon number squeezed state or a nonclassical state must satisfy the relation $\Delta n)^2 < \langle n\rangle$, where $n=a^\dagger a$ is the number operator for the generalized model. This implies that for a squeezed/nonclassical state the Mandel parameter $\mathcal{Q}$ \cite{Mandel}
\begin{equation} \label{MandelQ}
\mathcal{Q}=\frac{(\Delta n)^2}{\langle n\rangle}-1,
\end{equation}
must be negative, so that the corresponding distribution is sub-Poissonian. Clearly, $\mathcal{Q}=0$ corresponds to the Poissonian distribution and, thus, the number squeezing is absent. Similarly, the super-Poissonian case $\mathcal{Q}>0$ is also not important to us, since it does not identify any squeezing property. Note that, the same analysis is applicable to our generalized squeezed states, since they have been constructed in the Fock basis. However, for curiosity one may ask a question what happens if we generalize the definition of the Mandel parameter, say by defining the generalized number operator $N=A^\dagger A$. Firstly, it may be inappropriate in our case to deal with the generalized Mandel parameter, since we are effectively dealing in the Fock basis and not in the generalized basis. But, most importantly, even if we construct a generalized Mandel parameter as $Q_{\text{g}}=[(\Delta N)^2/\langle N\rangle]-1$, the Poissonian distribution does not correspond to the case of $Q_{\text{g}}=0$. This is because the Mandel parameter in the generalized coherent state basis takes the form $Q_{\text{g}}=\langle [A,A^\dagger]\rangle -1$, and it is not guaranteed that $[A,A^\dagger]=1$ is fulfilled in general. Therefore, the information of photon number squeezing can not be extracted from the $Q_{\text{g}}$ measured in the generalized coherent state basis. A similar kind of observation was made earlier by two of the authors of this article in different studies \cite{Dey,Zelaya}. However, a proper understanding of the generalization of the definition of Mandel parameter is yet underway. Nevertheless, for our purpose we may not require that analysis, or if it requires, we have to keep it as an open problem, which is beyond our scope in the present article. The behavior of the Mandel parameter for different values of the squeezing parameter for the trigonometric Rosen-Morse system is depicted in Fig.~\ref{fig2}. Notice that the Mandel parameter becomes negative in some region in all of the cases, including the case corresponding to the coherent states, $\xi=0$. This is not surprising since the generalized coherent states are sometimes known to be slightly nonclassical as discussed in the introduction. Nevertheless, upon increasing the squeezing parameter $\xi$, the nonclassicality increases accordingly as expected. Note that, the subpanel in Fig.~\ref{fig2} demonstrates the behavior of the same functions as in the main panel, but for the higher value of $\alpha$, where it is clearly visible that the line corresponding to $\xi=0.6$ moves below than that of $\xi=0.4$, thus, showing a consistent behavior.
\begin{figure*}
\centering \includegraphics[scale=0.32]{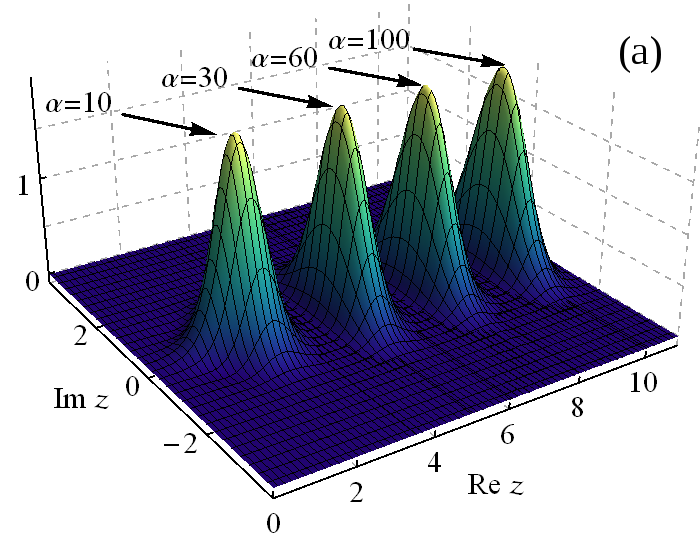} 
\includegraphics[scale=0.32]{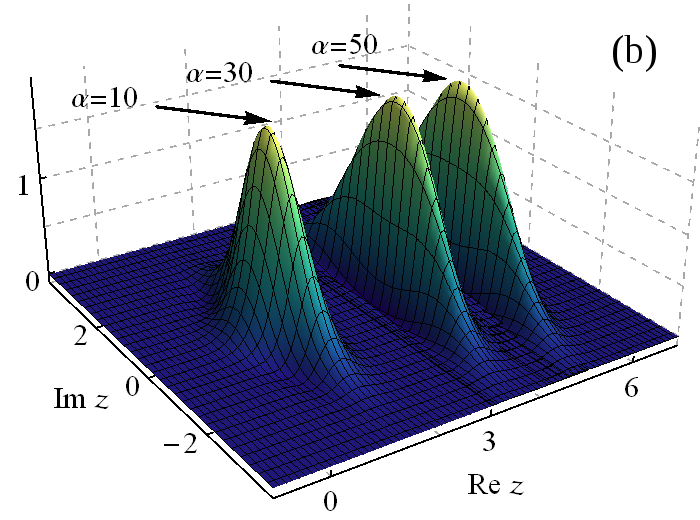}
\includegraphics[scale=0.32]{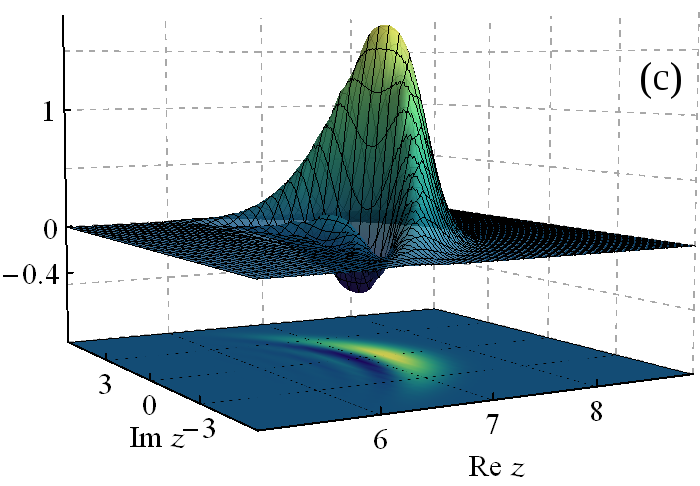} 
\includegraphics[scale=0.32]{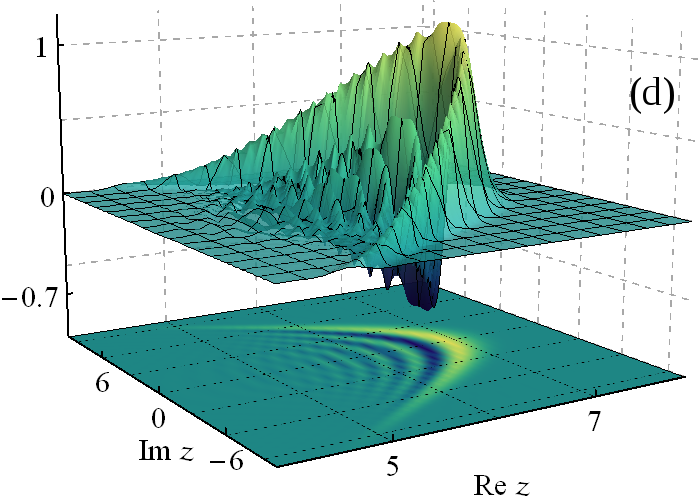}
\caption{Wigner functions for the Rosen-Morse squeezed states for (a) $\xi = 0$ (b) $\xi = 0.6$ (c) $\alpha=100,\xi = 0.8$ (d) $\alpha=100,\xi = 0.95$.}
\label{fig3}
\end{figure*}
\subsection{Wigner quasi-probability distribution}
The nonclassicality can also be described by the study of the Wigner quasi-probability distribution \cite{Wigner}, which is the quantum analogue of the phase space probability density. Thus, any negative region in the Wigner function indicates that the state possesses nonclassicality \cite{Wigner}. However, in order to study the Wigner function in the given scenario, we need to consider a construction for the most general density matrix, in terms of the Fock basis 
\begin{equation}
\rho=\displaystyle\sum_{m_1,m_2=0}^{\infty}\mathcal{C}_{m_1,m_2}|m_1\rangle\langle m_2|,
\end{equation}
with $\mathcal{C}_{m_1,m_2}=\langle m_1|\rho|m_2\rangle$ so that the corresponding Wigner function \cite{Wigner,Agarwal_Tara} becomes
\begin{alignat}{1} \label{WignerFn}
W(z)= & e^{2|z|^2}\displaystyle\sum_{m_1,m_2=0}^{\infty}\frac{\mathcal{C}_{m_1,m_2}}{\sqrt{m_1!m_2!}}\\
& \times\int\frac{d^2\beta}{\pi}e^{-|\beta|^2}(-\beta^\ast)^{m_1}(\beta)^{m_2} e^{2(z\beta^\ast-z^\ast\beta)}, \notag
\end{alignat}
where $z,\beta$ are the eigenvalues of the Glauber coherent states. Introducing a change of variable $\gamma=2z$ and, subsequently, using the identity $G(\gamma)=\int\frac{d^2\beta}{\pi}e^{-|\beta|^2} e^{\gamma\beta^\ast-\gamma^\ast\beta}=e^{-|\gamma|^2}$, the Wigner function \myref{WignerFn} can be rewritten in the following form
\begin{equation}\label{WignerFn1}
W(\gamma)=e^{|\gamma|^2/2}\displaystyle\sum_{m_1,m_2=0}^{\infty} \mathcal{C}_{m_1,m_2}\mathcal{F}_{m_1,m_2}(\gamma),
\end{equation} 
with $\mathcal{F}_{m_1,m_2}(\gamma)=\frac{(-1)^{m_1+m_2}}{\sqrt{m_1!m_2!}}\frac{\partial^{m_1+m_2}}{\partial\gamma^{m_1}\partial\gamma^{\ast m_2}}e^{-|\gamma|^2}$. By using the fact that the derivative of any analytic function $f(z,z^\ast)$ with respect to $z$ are independent of $z^\ast$ and vice-versa and, by employing the Rodrigues formula for the associated Laguerre polynomials, it can be shown that
\begin{alignat}{1}\label{FF}
&\mathcal{F}_{m_1,m_2}(z)=\\
&\left\{ \begin{array}{ll}
(-1)^{m_1}\sqrt{\frac{m_1!}{m_2!}}e^{-4|z|^2}(2z)^{m_2-m_1}L_{m_1}^{m_2-m_1}(4|z|^2), & m_2\geq m_1 \\ (-1)^{m_2}\sqrt{\frac{m_2!}{m_1!}}e^{-4|z|^2}(2z^\ast)^{m_1-m_2}L_{m_2}^{m_1-m_2}(4|z|^2), & m_2\leq m_1.\end{array}\right. \notag
\end{alignat} 
Here, we change the variable $z=\gamma/2$ again, so that we come back to the expression in terms of the original variable. Note that the Wigner function given by \myref{WignerFn1} along with \myref{FF} is completely general and can be applied to any quantum state irrespective of pure or mixed. However, it is slightly difficult to perform numerical computation with the expression we have, since it involves four sums. In what follows, we shall split the expression into some parts in order to reduce the computation time. Notice that the expression inside the double summation in \myref{WignerFn1} are the elements of an infinite dimensional matrix, therefore, we rewrite the Wigner function as $W(z)=\sum_{m_1,m_2=0}^{\infty}\Phi_{m_1,m_2}$, with $\Phi_{m_1,m_2}=e^{2|z|^2}\mathcal{C}_{m_1,m_2}\mathcal{F}_{m_1,m_2}(z)$. Thus, one can decompose the matrix $W(z)$ in terms of the diagonal $\Phi_{m,m}$, upper off-diagonal $\Phi_{m_1,m_2}$ and lower off-diagonal $\Phi_{m_2,m_1}$ elements as
\begin{equation}\label{Eq22}
W(z)=\displaystyle\sum_{m=0}^{\infty}\Phi_{m,m}+ \displaystyle\sum_{m_1=0}^{\infty}\displaystyle\sum_{m_2=m_1+1}^{\infty}\left(\Phi_{m_1,m_2}+\Phi_{m_2,m_1}\right).
\end{equation}
The second term can be written in terms of the upper off-diagonal elements only as
\begin{equation} \label{Eq24}
\Phi_{m_1,m_2}+\Phi_{m_2,m_1}=2\text{Re} \left[\Phi_{m_1,m_2}\right], ~~\forall ~m_2>m_1.
\end{equation} 
By replacing \myref{Eq24} in \myref{Eq22}, we obtain the final expression of the Wigner function, which reduces the computation time substantially. Let us now apply the scheme to the squeezed states of our system and compute the Wigner function numerically. The results of the computation for the trigonometric Rosen-Morse case are demonstrated in Fig.~\ref{fig3}. Interestingly, we observe the same behavior as in the case of the Mandel parameter, i.e. there are some very small negative regions in panel (a) in spite of the squeezing parameter $\xi$ being zero as per our expectation. Although the negative regions are not visible properly in the figure, because they are dominated by the positive peaks and, with certain rescaling they would have been visible. However, there is no way to test whether this negativity is due to the nonclassical property of the generalized coherent states, or due to the truncation of the infinite series. But, intuitively one may argue that it can cause because of the combined effects of these two. It should not occur only because of the truncation of the series, since in Fig.~\ref{fig2} we have noticed that the generalized coherent states may also possess little amount of nonclassicality. In any case, when we increase the squeezing parameter, as in panel (b), the negativity increases, as well as the distribution becomes squeezed in real $z$ axis. The rest of the figures as depicted in panel (c) and (d) are more interesting, since the negative peaks become stronger as we increase the squeezing parameter $\xi$ and, thus, move towards the more squeezed region.
\section{Concluding remarks}\label{sec5}      
We have proposed a scheme for the generalization of squeezed states along with an example of the trigonometric Rosen-Morse potential on which our protocol has been applied. Subsequently, we verify our results by studying the nonclassical properties of the system by utilizing several standard techniques; such as, quadrature squeezing, photon number squeezing and Wigner function. All of them agree that the states emerged out of the generalized system are always nonclassical and the degree of nonclassicality increases with the increase of the squeezing of the corresponding system. Although, the proposed generalization contains slightly complicated expression due to which all of the nonclassical properties have to be studied numerically, however, it is conceivable that the generalization appears with consistent results without any irregularity or shortcomings. Moreover, the generalization comes out with a relatively closed expression and, thus, it is legitimate that a further simplified version should follow up in near future subjecting to more intense investigation.

\vspace{0.5cm} \noindent \textbf{\large{Acknowledgements:}} K.Z. would like to thank V.H. and Centre de Recherches Math\'ematiques for kind hospitality, also acknowledges the support of CONACyT scholarship 45454 and the FRQNT international internship award 210974. S.D.~acknowledges the INSPIRE Faculty  research Grant (DST/INSPIRE/04/2016/001391) by the Department of Science and Technology, Govt.~of India. V.H.~acknowledges the support of research grants from CRSNG of Canada.    
\section*{Appendix}\label{sec6}
This section contains a detailed proof of the solution \myref{GenSqState} of the recurrence relation \myref{recurrence}. First, we substitute the solution \myref{GenSqState} into the recurrence relation \myref{recurrence}, so that we obtain
\begin{alignat}{1}
& \displaystyle\sum_{m=0}^{[\frac{n+1}{2}]}(-\xi)^m\alpha^{n-2m+1}g(n+1,m)\\
& -\displaystyle\sum_{m=0}^{[n/2]}(-\xi)^m\alpha^{n-2m+1}g(n,m) \notag\\
& +k(n)\displaystyle\sum_{m=0}^{[\frac{n-1}{2}]}(-1)^m\xi^{m+1}\alpha^{n-2m-1}g(n-1,m)=0, \notag
\end{alignat}
where $g(n,m)$ is given by \myref{gnm}. Subsequently, we collect the terms of same power of $\alpha$ and separate the even and odd powers of $n$, so that for even case we obtain
\begin{alignat}{1} \label{even}
& \xi\alpha^{2n-1}\left[k(2n)+g(2n,1)-g(2n+1,1)\right] \\
& +\displaystyle\sum_{m=1}^{n-1}(-\xi)^{m-1}\alpha^{2n-2m-1}\left[g(2n+1,m+1)\right. \notag \\
&\left. ~~~~ -g(2n,m+1)-k(2n)g(2n-1,m)\right]=0, \notag
\end{alignat}
and for the odd case it turns out to be
\begin{alignat}{1}\label{odd}
& \xi\alpha^{2n}\left[k(2n+1)+g(2n+1,1)-g(2n+2,1)\right] \\
& +(-\xi)^n\left[k(2n+1)g(2n,n)-g(2n+2,n+1)\right] \notag \\
& +\displaystyle\sum_{m=1}^{n-1}(-\xi)^{m-1}\alpha^{2n-2m}\left[g(2n+2,m+1)\right. \notag \\
& \left.~~~~ -g(2n+1,m+1)-k(2n+1)g(2n,m) \right]=0. \notag
\end{alignat}
As per the requirement, all the coefficients of $\alpha$ in both of the cases must vanish. Hence, we end up with some simple relations. For the even case \myref{even} results to
\begin{alignat}{1}
& g(2n+1,1)=g(2n,1)+k(2n), \label{id1}\\
& g(2n+1,m+1)=g(2n,m+1)+k(2n)g(2n-1,m), \label{id2}
\end{alignat} 
whereas for the odd case they become
\begin{alignat}{1}
& g(2n+2,n+1)=k(2n+1)g(2n,n), \label{id3}\\
& g(2n+2,1)=g(2n+1,1)+k(2n+1), \label{id4}\\
& g(2n+2,m+1)=g(2n+1,m+1) \label{id5}\\
&\qquad\qquad\qquad\qquad~~ +k(2n+1)g(2n,m). \notag
\end{alignat}
Now, \myref{id1}, \myref{id4} and \myref{id2}, \myref{id5} are the equivalent sets of equations, apart from a change of parameter. Therefore, effectively we are left with three identities \myref{id1}, \myref{id2} and \myref{id3} to prove, which are indeed straightforward by following the definition of $g(n,m)$ from \myref{gnm}.



\end{document}